\documentstyle[aps,epsfig,multicol]{revtex}

\begin{document}
\draft

\title{One-dimensional tight-binding models with correlated \\
diagonal and off-diagonal disorder}

\author{L.~Tessieri, F.~M.~Izrailev}

\address{
Instituto de F\'{\i}sica, Universidad Aut\'{o}noma de Puebla, \\
Apdo. Postal J-48, Puebla, Pue. 72570, M\'{e}xico}

\date{29th October 1999}

\maketitle

\begin{abstract}
We study localization properties of electronic states in one-dimensional
lattices with nearest-neighbour interaction. Both the site energies and
the hopping amplitudes are supposed to be of arbitrary form.
%No restriction is imposed
%on the form of the site potential and of the hopping amplitudes.
A few cases are considered in details. We discuss first the case in which
both the diagonal potential and the fluctuating part of the hopping
amplitudes are small. In this case we derive a general analytical expression
for the localization length, which depends on the pair correlators of the
diagonal and off-diagonal matrix elements.
The second case we investigate is that of strong uncorrelated disorder, for
which approximate analytical estimates are given and compared with numerical
data. Finally, we study the model with short-range correlations which
constitutes an extension with off-diagonal disorder of the random dimer model.
\end{abstract}


\begin{multicols}{2}
\section{Introduction}
The studies on the localization of electronic states in disordered solids
can be traced back to the seminal paper of P. W. Anderson~\cite{And58}
where the localization of quantum states was first discussed in connection
with transport properties of a random lattice.
This is a natural association, since the extended or localized nature of
the one-particle wave-functions in a disordered system plays a key role in
the determination of the metallic or insulating character of the system
itself.
As an outstanding example of the link between transport properties and
wave-function localization in random media one could cite, for instance,
the metal-insulator transition that occurs in disordered three-dimensional
samples upon increasing the randomness of the material.

The importance of the Anderson localization, however, far exceeds the
limited field of condensed matter physics. That is mainly due to the
fact that quantum interference is the fundamental mechanism responsible
for the electron localization in a random lattice; as a consequence,
analogous localization effects can take place in any phenomenon that
involves propagation of waves in a disordered medium~\cite{She95}. 
The same conceptual framework thus allows one to understand seemingly
heterogeneous phenomena as the localization of water waves~\cite{Lin86},
the coherent backscattering of photons~\cite{Alb85}, and the universal
conductance fluctuations typical of mesoscopic samples (see, for
instance, reference~\cite{A85,Lee85}).

The basic one-dimensional (1D) Anderson model has proved to be a valuable
tool to gain insight in the complex phenomenon of localization.
The model is appealing because it is extremely simple and yet retains
the capacity to provide a non-trivial description of the localization
process. Mathematically, the Anderson model is defined by the tridiagonal
Hamiltonian
\begin{equation}
\hat{H} = \sum_{n} \left[ \left| n\rangle 
\langle n-1 \right| + \left| n \rangle 
\langle n+1 \right| + \left| n \rangle \delta_{n} \langle n \right| \right]
\label{andmod}
\end{equation}
and disorder is introduced through the site energies $\delta_{n}$ which
are supposed to be {\em random} and {\em independent} variables (with
an appropriate distribution).

In spite of its undiscussed usefulness, the basic model~(\ref{andmod})
cannot account for many essential aspects of localization phenomena,
such as the metal-insulator transition.
The twofold reason of this shortcoming lies in the 1D nature
of the system~(\ref{andmod}) and in the statistical independence of the
random site energies $\delta_{n}$. 
As a consequence of these two key features, all eigenfunctions of the
Hamiltonian~(\ref{andmod}) turn out to be {\em localized}, even if the
disorder is infinitesimally small.

The key role played by dimensionality in the localization processes led at
first to the conclusion that only a 3D generalization of the Anderson
model~(\ref{andmod}) could reproduce such basic characteristics of disordered
solids as the existence of mobility edges or the metal-insulator transition.
The progress in the investigation of 1D systems, however, eventually led to
the realization that such models can exhibit a richer behaviour than it was
previously thought. In fact, it was discovered that even in one-dimensional
lattices there are classes of random potentials which allow for
{\em extended} states (see, e.~g., reference~\cite{Flo89}). In all these 1D
systems the disorder exhibits spatial correlations, in contrast to the
totally uncorrelated potential of the standard Anderson model. A
significant example of a system where the correlations of the site
potential can produce delocalized states is given by the so-called `random
dimer', which is characterized by a peculiar form of short-range
correlations of the potential~\cite{Dun90}.

The analysis of models with correlated disorder revived the interest in
1D chains and revealed the essential role of correlations.
One must observe, however, that the short-range correlations of the
random dimer and similar systems can give rise at best to
a {\em discrete} set of extended quantum states: in other words,
the localization length diverges only for discrete values of the energy.
The situation may change when long-range correlations come into play, as
it was recently discovered in~\cite{Izr99}, where the general case of diagonal
disorder with arbitrary correlations was considered and a direct relation
between the localization length and the potential pair correlators was
established. Using this relation it was shown how to reconstruct a
site-potential giving rise to any given form of energy dependence of the
localization length~\cite{Izr99}. A particular conclusion is that even 1D
random lattices can possess a {\em continuum} of extended states (and
mobility edges), provided the disorder exhibits appropriate long-range
correlations.

In the present work we follow an approach similar to that of
reference~\cite{Izr99} by extending the study to a general case of 1D
Anderson model with any kind of diagonal disorder and off-diagonal
nearest-neighbour interaction.

This work is organized as follows. The rest of the introduction is devoted
to the definition of the model under study. In Section~\ref{wcd} we analyse
the case in which the site-energies and the site-dependent part of the hopping
amplitudes are small. Although we focus our discussion on the case of random
site- and hopping energies, the validity of our results extends to any
non-random case as well.
In Section~\ref{sud} we consider the case of strong and uncorrelated
disorder, deriving an approximate expression for the localization length
and comparing it with the numerical data.
In Section~\ref{dwsrc} we investigate the case (reminiscent of the random
dimer model) in which the disorder (both diagonal and off-diagonal) has
arbitrary intensity and exhibits peculiar short-range correlations.
The concluding remarks are exposed in Section~\ref{cr}.

\subsection{Definition of the model}

We study the localization properties of the following one-dimensional
tight-binding Hamiltonian
\begin{eqnarray}
\hat{H} & = & \sum_{n} \left[ \left| n\rangle \gamma_{n-1}
\langle n-1 \right| + \left| n \rangle \gamma_{n+1}
\langle n+1 \right| \right.\nonumber \\
& & \left. + \left| n \rangle \delta_{n} \langle n \right| \right]
\label{ham}
\end{eqnarray}
%\begin{equation}
%\hat{H} = \sum_{n} \left[ \left| n\rangle \gamma_{n-1}
%\langle n-1 \right| + \left| n \rangle \gamma_{n+1}
%\langle n+1 \right| + \left| n \rangle \delta_{n} \langle n
%\right| \right]
%\label{ham}
%\end{equation}
where the site energies $\delta_{n}$ and the off-diagonal elements
$\gamma_{n}$ are arbitrary real variables. From the physical point of
view, the Hamiltonian~(\ref{ham}) describes an `electron' moving in a
discrete 1D lattice. The energies $\delta_{n}$ measure the strength of
the bond of the electron to the `atoms' of the chain, while the hopping
terms $\gamma_{n}$ represent the probability amplitudes for an electron
localized in a single site to jump to the nearest atoms to the right or to
the left.

Notice that in our model the transition amplitudes
\begin{displaymath}
A_{n \rightarrow n \pm 1} = \langle n \pm 1 | H | n \rangle = \gamma_{n}
\end{displaymath}
depend only on the initial state of the electron and not on the final one.
This physical feature is connected to the non-Hermitian character of the
operator~(\ref{ham}).
With regard to this point, we stress that, although the operator~(\ref{ham})
is not Hermitian, its eigenvalues are real: this allows one to consider it
as a physically sound Hamiltonian.

In order to show that the eigenvalues of $\hat{H}$ are real, let $E$ be
an eigenvalue of the operator~(\ref{ham}) and $|\psi_{E} \rangle$ the
corresponding eigenvector
\begin{equation}
\hat{H} \left| \psi_{E} \right. \rangle =
E \left| \psi_{E} \right. \rangle .
\label{dim1}
\end{equation}
Let us now consider the operator $\hat{H}^{T}$, i.e. the transpose
of $\hat{H}$ and let us denote with $|\phi_{E} \rangle$ the eigenvector of
$\hat{H}^{T}$ to the eigenvalue $E$, so that one has
\begin{equation}
\hat{H}^{T} \left| \phi_{E} \right. \rangle =
E \left| \phi_{E} \right. \rangle .
\label{dim2}
\end{equation}
Notice that if $E$ belongs to the spectrum of $\hat{H}$, then the same
value is also an eigenvalue of $\hat{H}^{T}$, since a matrix and its
transpose share the same spectrum.
Multiplying equation~(\ref{dim2}) by $\langle \psi_{E} |$ one can obtain
\begin{eqnarray}
E \langle \left. \psi_{E} \right| \phi_{E} \rangle =
\langle \left. \psi_{E} \right| \hat{H}^{T} \left| \phi_{E} \right.
\rangle = \nonumber \\
\langle \hat{H} \psi_{E} \left| \right. \phi_{E} \rangle =
E^{*} \langle \psi_{E} \left| \right. \phi_{E} \rangle
\label{dim3}
\end{eqnarray}
where we have used equation~(\ref{dim1}) and the fact that $\hat{H}$ is
a real matrix (and therefore its transpose coincides with its adjoint).
Equation~(\ref{dim3}) implies that $E = E^{*}$, unless one has
$\langle \psi_{E} \left| \right. \phi_{E} \rangle = 0$. One can rule this
last possibility out, however, due to the specific structure of the
operator~(\ref{ham}).
Indeed, for the Hamiltonian~(\ref{ham}) one has that
\begin{equation}
\hat{H}^{T} = \hat{A}^{-1} \hat{H} \hat{A}
\label{dim4}
\end{equation}
where $\hat{A} = \sum_{n} \left| n \right. \rangle \frac{1}{\gamma_{n}}
\langle \left. n \right|$. Relation~(\ref{dim4}) between $\hat{H}$ and
$\hat{H}^{T}$ implies that $| \phi_{E} \rangle =
\hat{A}^{-1} |\psi_{E} \rangle$, and this allows one to exclude the
possibility that $|\psi_{E} \rangle$ and $| \phi_{E} \rangle$ are
orthogonal. This leads to the conclusion that $E = E^{*}$, i.e. that the
Hamiltonian~(\ref{ham}) has real eigenvalues.

Below, we consider the stationary Schr\"{o}dinger
equation $\hat{H} \left| \psi \right. \rangle = E \left| \psi \right.
\rangle$, with eigenvectors
\begin{displaymath}
\left| \psi \right. \rangle = \sum_{n} \psi_{n} \left| n \right. \rangle .
\end{displaymath}
Using the explicit form~(\ref{ham}) of the Hamiltonian, one can easily
see that the Schr\"{o}dinger equation for the amplitudes $\psi_{n}$
takes the form
\begin{equation}
\gamma_{n+1} \psi_{n+1} + \gamma_{n-1}  \psi_{n-1} =
\left( E - \delta_{n} \right) \psi_{n} .
\label{schr1}
\end{equation}
It is convenient to introduce the new variables
\begin{equation}
\phi_{n} = \gamma_{n} \psi_{n}
\label{phi}
\end{equation}
and
\begin{equation}
\xi_{n} \left( E \right) = \frac{E - \delta_{n}}{\gamma_{n}} ,
\label{xi}
\end{equation}
so that equation~(\ref{schr1}) can be cast in the simpler form
\begin{equation}
\phi_{n+1} + \phi_{n-1} = \xi_{n} \phi_{n} .
\label{schr2}
\end{equation}
The introduction of the amplitudes~(\ref{phi}) and of the
site-energies~(\ref{xi}) thus formally allows one to reduce the problem of
determining the eigenstates of model~(\ref{ham}) to the equivalent problem
of studying the zero-energy eigenstate of the tridiagonal
model~(\ref{schr2}) with diagonal-only disorder.
According to this interpretation, for any energy value $E$ in the original
Schr\"{o}dinger equation there is a corresponding realization
$\left\{ \xi_{n} \left( E \right) \right\}$ in~(\ref{schr2}).

Having thus defined our problem, we proceed to solve it for three distinct
physical cases: the limiting case of weak disorder (with arbitrary
correlations), the opposite case of strong (uncorrelated) disorder and the
case of disorder of arbitrary strength with short-range correlations
(`$N$-mer model').
In the following, we will consider off-diagonal terms of the form
\begin{equation}
\gamma_{n} = 1 + \epsilon_{n} .
\label{gamma}
\end{equation}
The reason for representing the energies $\gamma_{n}$ in the
form~(\ref{gamma}) is that in this way we separate two physically distinct
contributions to the off-diagonal matrix elements: the first one originates
from the kinetic energy term of the Hamiltonian, while the second one,
$\epsilon_{n}$, represents the fluctuating part of the interaction
between nearest neighbours. The latter contribution is the source
of randomness for the hopping energies $\gamma_{n}$ of our
model~(\ref{ham}).

\section{Weak correlated disorder}
\label{wcd}

\subsection{Localization length}

In this Section we consider the general case of {\em weak} disorder:
\begin{equation}
\begin{array}{lcl}
|\epsilon_{n}| \ll 1 & \mbox{and} & |\delta_{n}| \ll 1
\end{array}
\label{small}
\end{equation}
for {\em any} kind of site-energies $\delta_{n}$ and interaction
$\epsilon_{n}$.

Notice that, since the randomness of the hopping terms $\gamma_{n}$ comes
only from the interaction energies $\epsilon_{n}$, the condition
of weak disorder for the off-diagonal terms is given by the first of the
relations~(\ref{small}) (and not by the condition $\gamma_{n} \ll 1$).
Under the conditions~(\ref{small}) one has
\begin{displaymath}
\xi_{n} \simeq E - \left( E \epsilon_{n} + \delta_{n} \right)
\end{displaymath}
so that equation~(\ref{schr2}) can be approximated by
\begin{equation}
\phi_{n+1} + \phi_{n-1} = \left( E - E \epsilon_{n} - \delta_{n} \right)
\phi_{n} .
\label{schr3}
\end{equation}

\subsubsection{Classical representation of the quantum model}

An effective way~\cite{Izr95} to study the quantum model~(\ref{schr3})
consists in representing it in terms of the classical two-dimensional
Hamiltonian map
\begin{equation}
\left\{
\begin{array}{ccc}
p_{n+1} & = & \left( p_{n} + A_{n} x_{n} \right) \cos \mu + x_{n}
\sin \mu \\
x_{n+1} & = & -\left( p_{n} + A_{n} x_{n} \right) \sin \mu + x_{n}
\cos \mu 
\end{array}
\right.
\label{map}
\end{equation}
with
\begin{equation}
E = 2 \cos \mu
\label{energy}
\end{equation}
and
\begin{equation}
A_{n} = \frac{\delta_{n} + 2 \epsilon_{n} \cos \mu}{\sin \mu} .
\label{hit}
\end{equation}
The map~(\ref{map}) describes the behaviour of a harmonic oscillator
subjected to periodic delta kicks of amplitude $A_{n}$. The map can in fact
be derived by integrating over a period $T=1$ the equations of motion of
the kicked oscillator defined by the Hamiltonian
\begin{displaymath}
H = \mu \left( \frac{p^{2}}{2} + \frac{x^{2}}{2} \right) + \frac{x^{2}}{2}
\sum_{n=-\infty}^{\infty} A_{n} \delta \left( t - n \right) .
\end{displaymath}

The complete equivalence of the quantum system~(\ref{schr3}) with the classical
one defined by the map~(\ref{map}) can be easily proved by eliminating the $p$
variable from equations~(\ref{map}). In this way, one obtains the relation
\begin{displaymath}
x_{n+1} + x_{n-1} =\left( 2 \cos \mu - A_{n} \sin \mu \right) x_{n}
\end{displaymath}
which coincides with equation~(\ref{schr3}) if one identifies the variable
$x_{n}$ with the site amplitude $\phi_{n}$ and if the parameters of the
quantum model~(\ref{schr3}) and those of the classical map~(\ref{map}) are
related by the equations~(\ref{energy}) and~(\ref{hit}).
Therefore it becomes possible to analyse the solutions of
equation~(\ref{schr3}) in terms of `trajectories' in the phase-space of
the map~(\ref{map}). In such an approach, localized quantum states
correspond to unbounded trajectories in the classical phase-space,
while extended quantum states are represented by bounded trajectories
(see details in~\cite{Izr95}).

It is convenient to express the map~(\ref{map}) in polar coordinates
$(r,\theta)$ introduced via standard relations $p = r \cos \theta$
and $x = r \sin \theta$. The substitution in equation~(\ref{map}) gives
\begin{displaymath}
\left\{
\begin{array}{ccc}
\cos \theta_{n+1} & = & \frac{1}{D_{n}} \left[ \cos \left( \theta_{n} -
\mu \right) + A_{n} \sin \theta_{n} \cos \mu \right]  \\
\sin \theta_{n+1} & = & \frac{1}{D_{n}} \left[ \sin \left( \theta_{n} -
\mu \right) - A_{n} \sin \theta_{n} \sin \mu \right]
\end{array}
\right.
\end{displaymath}
where
\begin{displaymath}
D_{n} = \frac{r_{n+1}}{r_{n}} = \sqrt{1 + A_{n} \sin \left( 2 \theta_{n}
\right) + A_{n}^{2} \sin^{2} \theta_{n} }.
\end{displaymath}
Using this approach, the inverse localization length $l^{-1}$ (or
Lyapunov exponent $\lambda$) can be expressed as
\begin{eqnarray}
l^{-1} & = & \lambda = \lim_{N \rightarrow \infty} \frac{1}{N}
\sum_{n=1}^{N} \log \frac{x_{n+1}}{x_{n}} \nonumber \\
 & = & \lim_{N \rightarrow \infty} \frac{1}{N} \sum_{n=1}^{N}
\log \left( \frac{r_{n+1} \sin \theta_{n+1}}{r_{n} \sin \theta_{n}} \right)
\label{lyap1}
\end{eqnarray}
Except that at the band edge (i.e. for $|E| \rightarrow 2$),
expression~(\ref{lyap1}) can be safely reduced to
\begin{equation}
\lambda =  \lim_{N \rightarrow \infty} \frac{1}{N} \sum_{n=1}^{N}
\log \left( \frac{r_{n+1}}{r_{n}} \right)
= \lim_{N \rightarrow \infty} \frac{1}{N} \sum_{n=1}^{N} \log D_{n}
\label{lyap2}
\end{equation}
(see \cite{Izr98} for details).

For weak disorder, the logarithm in~(\ref{lyap2}) can be expanded and
to the second order of perturbation theory one gets
\begin{equation}
\lambda = \frac{1}{8} \langle A_{n}^{2} \rangle + \frac{1}{2} \langle
A_{n} \sin \left( 2 \theta_{n} \right) \rangle
\label{lyap3}
\end{equation}
where the symbol $ \langle \cdots \rangle$ stands for the `time'-average,
$\langle x_{n} \rangle = \lim_{N \rightarrow \infty} \frac{1}{N}
\sum_{n=1}^{N} x_{n}$.
%\begin{displaymath}
%\langle x_{n} \rangle = \lim_{N \rightarrow \infty} \frac{1}{N}
%\sum_{n=1}^{N} x_{n}  .
%\end{displaymath}

\subsubsection{Computation of the noise-angle correlator}
To compute the correlator $\langle A_{n} \sin \left( 2 \theta_{n} \right)
\rangle$  with second-order accuracy we follow the approach of~\cite{Izr99}.
We define the noise-angle $a_{k}$ and the noise-noise $q_{k}$ correlators
as follows,
\begin{equation}
a_{k} = - \frac{2i}{\sigma_{A}^{2}} e^{2i \mu} \langle A_{n}
e^{2i \theta_{n-k}} \rangle
\label{nacorr}
\end{equation}
and
\begin{eqnarray}
q_{k} & = & \frac{1}{\sigma_{A}^{2}} \langle A_{n} A_{n-k} \rangle
\nonumber \\
 & = &
\frac{E^{2} \langle \epsilon_{n} \epsilon_{n-k} \rangle +2E \langle
\epsilon_{n} \delta_{n-k} \rangle + \langle \delta_{n} \delta_{n-k}
\rangle}{E^{2} \langle \epsilon_{n}^{2} \rangle +2E \langle \epsilon_{n}
\delta_{n} \rangle + \langle \delta_{n}^{2} \rangle}
\label{nncorr}
\end{eqnarray}
where $\sigma_{A}^{2}$ is the `noise'-variance
\begin{displaymath}
\sigma_{A}^{2} = \langle A_{n}^{2} \rangle =
\frac{E^{2} \langle \epsilon_{n}^{2} \rangle + 2 E \langle \epsilon_{n}
\delta_{n} \rangle + \langle \delta_{n}^{2} \rangle}{1 - E^{2}/4}.
\end{displaymath}
The terms~(\ref{nacorr}) measure the `temporal' correlations of the
noise $A_{n}$ with the angle variable $\theta_{n}$, while
equation~(\ref{nncorr}) defines the normalized autocorrelation function
for the noise variable~(\ref{hit}).
Notice that, since the noise strength~(\ref{hit}) is a function both of
random terms $\epsilon_{n}$ and $\delta_{n}$ and of the energy $E = 2
\cos \mu$, the noise-noise correlator~(\ref{nncorr}) depends on the energy
$E$ as well as on the (spatial) correlators of the random variables
$\delta_{n}$ and $\epsilon_{n}$.

From the definitions~(\ref{nacorr}), (\ref{nncorr}) it follows that
\begin{equation}
\langle A_{n} \sin \left( 2 \theta_{n} \right) \rangle = \mbox{Re} \left(
\frac{\sigma_{A}^{2}}{2} e^{-2i \mu} a_{0} \right) .
\label{rel1}
\end{equation}
As in reference~\cite{Izr99} it can be shown that the correlators $a_{k}$
satisfy the relations
\begin{displaymath}
a_{k-1} = e^{-2i \mu} a_{k} + q_{k}, \; \; \; \; \mbox{for } k=1,2,3,\ldots
\end{displaymath}
which in turn imply that
\begin{equation}
a_{0} = \sum_{k=1}^{\infty} q_{k} e^{-2i \mu (k-1)} .
\label{rel2}
\end{equation}
Substituting~(\ref{rel2}) into~(\ref{rel1}), one can write
\begin{displaymath}
\langle A_{n} \sin \left( 2 \theta_{n} \right) \rangle =
\frac{\sigma_{A}^{2}}{2} \sum_{k=1}^{\infty} q_{k} \cos \left( 2 \mu k
\right)
\end{displaymath}
so that the Lyapunov exponent~(\ref{lyap3}) takes the form
\begin{equation}
\lambda = \lambda_{0} \varphi \left( \mu \right)
\label{lyap4}
\end{equation}
where
\begin{equation}
\lambda_{0} \left( E \right) = \frac{\sigma_{A}^{2}}{8} =
\frac{\langle \delta_{n}^{2} \rangle + 2 E \langle \epsilon_{n}
\delta_{n} \rangle + E^{2} \langle \epsilon_{n}^{2} \rangle}
{8 \left( 1-E^{2}/4 \right)}
\label{notcorr}
\end{equation}
and
\begin{equation}
\varphi \left( \mu \right) = 1 + 2 \sum_{k=1}^{\infty}
q_{k} \left( \mu \right) \cos \left( 2 \mu k \right)
\label{normlyap}
\end{equation}
while the parameter $\mu$ is related to the energy $E$ through
equation~(\ref{energy}).

\subsubsection{Discussion of the results}

Expression~(\ref{lyap4}), together with equations~(\ref{notcorr})
and~(\ref{normlyap}), gives the localization length as a function of
the energy and of the diagonal and off-diagonal potential correlators.
Notice that this is a very general result, because its validity rests
exclusively on the weak disorder assumption~(\ref{small}) and does
{\em not} depend on the particular form of the variables $\delta_{n}$
and $\epsilon_{n}$.
Equation~(\ref{lyap4}) can therefore be applied to a broad variety of
specific problems.

A simple examination of formula~(\ref{lyap4}) also reveals that
the inverse localization length $\lambda$ is the product of two factors,
$\lambda_{0}$ and $\varphi \left( \mu \right)$, defined respectively by
equation~(\ref{notcorr}) and~(\ref{normlyap}). This factorization is
physically meaningful, because $\lambda_{0}$ represents the Lyapunov
exponent for the case of totally uncorrelated disorder (i.e. when
$q_{k} = 0$ for $k \ge 1$), while the function $\varphi \left( \mu \right)$
describes the correction introduced by the spatial correlations both in
diagonal and off-diagonal terms.

Let us now analyse separately the two factors~(\ref{notcorr})
and~(\ref{normlyap}). As was indicated, the first one gives the inverse
localization length when no correlations exist among the matrix elements
of the Hamiltonian~(\ref{ham}). This system corresponds to a simple
generalization of the Anderson model, in which not only the site energies
but also nearest-neighbour hopping energies are independent random
variables. From this point of view, the expression~(\ref{notcorr}) can be
seen as an extension of the known formula for the standard Anderson model,
\begin{displaymath}
\lambda = \frac{W^{2}}{96 \left( 1 - E^{2}/4 \right)} .
\end{displaymath}
Indeed the inverse localization length~(\ref{notcorr}) reduces to this
form if one assumes that there is no off-diagonal disorder (i.e. $\langle
\epsilon_{n}^{2} \rangle = 0$ and $\langle \epsilon_{n} \delta_{n} \rangle
= 0$) and that the variance of the site energies is the one
fixed by the Anderson distribution (i.e. $\langle \delta_{n}^{2} \rangle =
W^{2}/12$).
Expression~(\ref{notcorr}), however, also shows that the introduction of
off-diagonal randomness modifies in a non-trivial way the energy dependence
of the Lyapunov exponent with respect to the standard diagonal case.
In particular, extended states can arise for specific values of the energy:
if, for instance, the diagonal and off-diagonal disorder are identical
($\epsilon_{n} = \delta_{n}$), the localization length diverges for $E=-1$.
We discuss this feature in more detail in the next Subsection.

The factor~(\ref{normlyap}) accounts for the modifications generated
by the correlations both in the diagonal and off-diagonal terms. One
should stress that this factor depends both on the correlations and
energy. This is a key feature, because it implies that, as an effect of
the correlations, the sample can become more or less transparent for
specific values of the energy. As can be seen from equation~(\ref{lyap4}),
if $\varphi \left( \mu \right) > 1$, the Lyapunov exponent is increased
with respect to the uncorrelated case, meaning that correlations
make the sample more opaque.
On the contrary, when $\varphi \left( \mu \right) < 1$, the chain becomes
more transparent. Thus, if the function~(\ref{normlyap}) goes to zero in a
certain energy range, the states of the system will be extended inside that
region and localized outside of it: this shows how mobility edges can arise
even in 1D systems, provided that the potential has the proper correlations.

Once the statistical properties of the potential are known,
formulae~(\ref{lyap4}), (\ref{notcorr}) and~(\ref{normlyap}) exactly
determine the Lyapunov exponent as a function of the energy. They can be
used, however, also to solve the inverse problem, i.e. to specify what
the correlations of the disorder must be for the Lyapunov exponent to
exhibit a given energy dependence.
This is a problem of practical interest because of the relevance of its
potential technological applications: knowing what particular disorder
generates a specific energy dependence of the Lyapunov exponent might
lead to the construction of superlattices with the required
transport properties.

Let us now write explicit relations between statistical properties of
the disorder and the Lyapunov exponent for the case when the diagonal
disorder is independent from the off-diagonal one ( $\langle \epsilon_{n}
\delta_{n-k} \rangle = 0$ ).
For this we write the function $\varphi \left( \mu \right)$ in Fourier
series,
\begin{equation}
\varphi \left( \mu \right) = \sum_{k = - \infty}^{+ \infty} \varphi_{k}
\; e^{i 2 \mu k}
\label{fourser}
\end{equation}
where the expansion coefficients are defined by the relation
\begin{displaymath}
\varphi_{k} = \frac{1}{\pi} \int_{- \frac{\pi}{2}}^{\frac{\pi}{2}}
\varphi \left( \mu \right) e^{-i 2 \mu k} \; d \mu .
\end{displaymath}
Inserting~(\ref{fourser}) and~(\ref{nncorr}) in equation~(\ref{normlyap})
one obtains the relations
\begin{eqnarray}
\langle \delta_{n}^{2} \rangle \varphi_{k} & + & \langle \epsilon_{n}^{2}
\rangle \left( \varphi_{k-1} + 2 \varphi_{k} + \varphi_{k+1} \right)
= \langle \delta_{n} \delta_{n-k} \rangle 
\nonumber \\
 & + & \langle \epsilon_{n}
\epsilon_{n-k-1} \rangle + 2 \langle \epsilon_{n} \epsilon_{n-k} \rangle
+ \langle \epsilon_{n} \epsilon_{n-k+1} \rangle
\label{invers}
\end{eqnarray}
which constitute the explicit link between the Fourier components of
$\varphi \left( \mu \right)$ and the disorder correlators
$\langle \delta_{n} \delta_{n-k} \rangle$ and $\langle \epsilon_{n}
\epsilon_{n-k} \rangle$.
Once the Lyapunov exponent (and hence the factor $\varphi \left( \mu
\right)$) is known, the relations~(\ref{invers}) set a constraint
on the statistical properties of $\delta_{n}$ and $\epsilon_{n}$.
Notice that, in contrast to the purely diagonal case discussed
in~\cite{Izr99}, here the interplay of diagonal and off-diagonal disorder
does not allow a complete specification of the correlators from the
knowledge of the localization length.

The relations~(\ref{invers}) admit a particularly simple solution for
the case in which the diagonal and off-diagonal disorder are
statistically independent (i.e. $\langle \epsilon_{n} \delta_{n-k}
\rangle = 0$), but share the same statistical properties (so that their
autocorrelation functions are equal: $\langle \delta_{n} \delta_{n-k}
\rangle = \langle \epsilon_{n} \epsilon_{n-k} \rangle$).
In this peculiar case, the relations~(\ref{invers}) {\em do} determine
the disorder correlators: indeed, one has
\begin{displaymath}
\frac{\langle \delta_{n} \delta_{n-k} \rangle}{\langle \delta_{n}^{2}
\rangle} = \frac{ \langle \epsilon_{n} \epsilon_{n-k} \rangle}
{\langle \epsilon_{n}^{2} \rangle} = \varphi_{k} .
\end{displaymath}
This result can be understood from the observation that in this
special case the noise-noise correlators~(\ref{nncorr}) take the form
$q_{k} = \langle \delta_{n} \delta_{n-k} \rangle / \langle \delta_{n}^{2}
\rangle$ and no longer depend on the energy, so that they can be interpreted
as Fourier coefficients of the series~(\ref{normlyap}). 

\subsection{Numerical data}
We checked numerically the validity of~(\ref{lyap4}) using the classical
map approach for the computation of the Lyapunov exponent.
We consider the map
\begin{equation}
\left( \begin{array}{c} P_{n+1} \\ X_{n+1} \end{array} \right) =
{\bf M}_{n}
\left( \begin{array}{c} P_{n} \\ X_{n} \end{array} \right) 
\label{map2}
\end{equation}
with
\begin{equation}
{\bf M}_{n} = \left( \begin{array}{cc}
\cos \mu & A_{n} \cos \mu + \gamma_{n} \sin \mu \\
- \frac{1}{\gamma_{n+1}} \sin \mu & \frac{1}{\gamma_{n+1}}
\left( \gamma_{n} \cos \mu -  A_{n} \sin \mu  \right)
\end{array} \right)
\label{mapmatrix}
\end{equation}
which is exactly equivalent to the original Schr\"{o}dinger
equation~(\ref{schr1}), provided the parameters $E$ and $A_{n}$ of the
matrix~(\ref{mapmatrix}) are defined by the relations~(\ref{energy})
and~(\ref{hit}).
This means that, if~(\ref{energy}) and~(\ref{hit}) are fulfilled, then the
values of the $X$ coordinate at the time steps $n-1$, $n$ and $n+1$ obey the
same relation expressed by~(\ref{schr1}) for the probability amplitudes
$\psi_{n}$.

It should be pointed out that the determinant of the single
time-step map~(\ref{mapmatrix}) is not unitary,
$\det {\bf M}_{n} = \gamma_{n}/\gamma_{n+1}$
%\begin{displaymath}
%\det {\bf M}_{n} = \frac{\gamma_{n}}{\gamma_{n+1}}
%\end{displaymath}
and, therefore, the determinant oscillates around the unit value as a
function of $n$. This does not mean, however, that the map~(\ref{map2})
does not conserve the total flux. Indeed, the disordered sample represented
by the Hamiltonian~(\ref{ham}) cannot actually be infinite; therefore some
kind of boundary conditions have to be imposed.
A typical boundary condition is the periodic one: in this case one assumes
that the disordered sample is a closed chain made up of $N$ `atoms' so that
the site- and hopping-energies obey the equations:
$\delta_{N+1} = \delta_{1}$ and $\gamma_{N+1} = \gamma_{1}$.
Consequently, one has
\begin{equation}
\prod_{k=1}^{N} \det {\bf M}_{k} = \frac{\gamma_{1}}{\gamma_{2}} \cdot
\frac{\gamma_{2}}{\gamma_{3}} \ldots \frac{\gamma_{N-1}}{\gamma_{N}} \cdot
\frac{\gamma_{N}}{\gamma_{N+1}} = \frac{\gamma_{1}}{\gamma_{N+1}} = 1
\label{totdet}
\end{equation}
so that the total determinant across the sample is exactly equal to one.
A more physical boundary condition arises if the disordered sample is
embedded between two perfect leads.
This physical condition translates in the mathematical requirements
$\delta_{n}=0$ and $\gamma_{n} = 1$ for $n > N$ or $n \leq 1$, which, in turn,
lead again to the condition~(\ref{totdet}) for the total determinant of
the map through the whole sample.

\begin{figure}[htb]
\begin{center}
\narrowtext
\epsfig{file=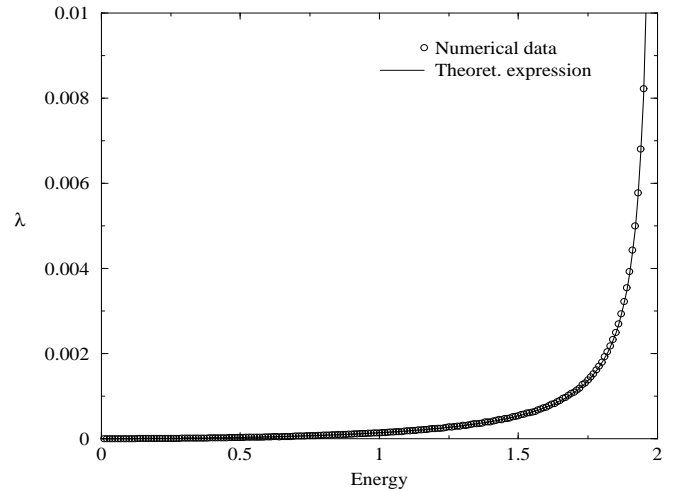,width=3.4in,height=2.6in}
\caption{Lyapunov exponent vs. energy.
The figure refers to the case $\alpha = 0$ and $W=0.1$.}
\label{weaal0}
\end{center}
\end{figure}

\begin{figure}[htb]
\begin{center}
\epsfig{file=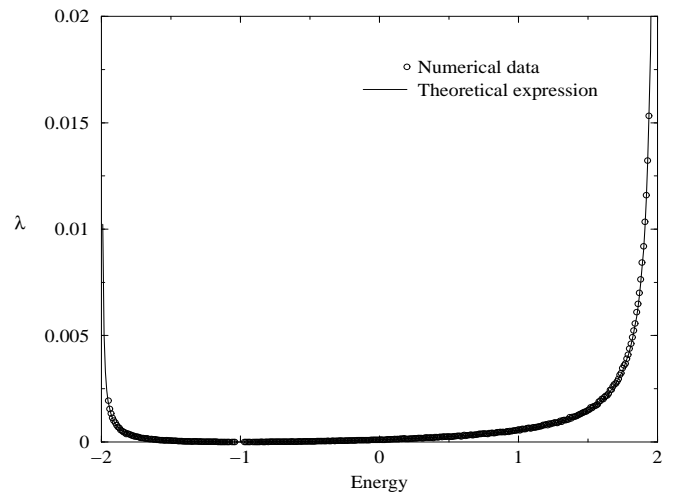,width=3.4in,height=2.6in}
\caption{Lyapunov exponent vs. energy. The figure refers to the case
$\alpha = 1$ and $W=0.1$}
\label{weaal1}
\end{center}
\end{figure}

Let us now see how the map~(\ref{map2}) can be effectively used for
the computation of the Lyapunov exponent. As was shown above, it is
convenient to introduce the polar coordinates $X = r \sin \theta$ and
$Y = r \cos \theta$ and cast the map~(\ref{map2}) in the form
\begin{equation}
\left\{ \begin{array}{ccl}
\cos \theta_{n+1} & = & \frac{r_{n}}{r_{n+1}} \left[ \cos \left( \theta_{n}
- \mu \right) \right. \\
 & & \left. + \left( A_{n} \cos \mu + \epsilon_{n} \sin \mu \right)
\sin \theta_{n} \right] \\
\sin \theta_{n+1} & = & \frac{r_{n}}{r_{n+1}} \frac{1}{1+\epsilon_{n+1}}
\left[ \sin \left( \theta_{n} - \mu \right) \right. \\
 & & \left. + \left( A_{n} \sin \mu + \epsilon_{n} \cos \mu \right)
\sin \theta_{n} \right]
\end{array}
\right.
\label{polmap}
\end{equation}
which allows an easy computation of the expression~(\ref{lyap2}) for the
Lyapunov exponent.
In this way, we tested the validity of the formula~(\ref{lyap4}) for the
special case in which the hopping energies $\epsilon_{n}$ are
{\em independent} random variables with a common distribution, while the
site-energies are defined by the relation
\begin{equation}
\begin{array}{ccc}
\delta_{n} = \alpha \epsilon_{n} & \mbox{ with } & \left| \alpha \right|
< 2 .
\end{array}
\label{diadis}
\end{equation}
Taking into account Eqs.~(\ref{diadis}) and~(\ref{gamma}), the
Schr\"{o}dinger equation~(\ref{schr1}) can be written as
\begin{equation}
\left( 1 + \epsilon_{n+1} \right) \psi_{n+1} + \left( 1 + \epsilon_{n-1}
\right) \psi_{n-1} = \left( E - \alpha \epsilon_{n} \right) \psi_{n} .
\label{schr4}
\end{equation}
The box distribution
\begin{equation}
p \left( \epsilon_{n} \right) = \frac{1}{W} \theta \left( \frac{W}{2} -
\left| \epsilon_{n} \right| \right)
\label{boxdis}
\end{equation}
of width $W$ was chosen for the variables $\epsilon_{n}$.
For this particular model the general expression~(\ref{lyap4}) reduces
to the following form;
\begin{equation}
\lambda = \frac{W^{2}}{96} \frac{{\left( E + \alpha \right)}^{2}}
{\left( 1 - E^{2}/4 \right)} .
\label{lyap5}
\end{equation}
This formula, among other things, shows that for the above-defined model
the state corresponding to the energy value $E = - \alpha$ is always
extended, regardless of the noise strength fixed by the parameter $W$.
In particular, when $\alpha = 0$ (i.e. there is no diagonal disorder),
the model~(\ref{schr4}) exhibits an extended state at the band centre, i.e.
for $E=0$. The value $\alpha = 1$ corresponds to the case, cited in the
previous Subsection, in which the binding and hopping energies are equal
($\epsilon_{n} = \delta_{n}$) and one has a transparent state for $E=-1$.
The comparison with numerical data revealed an excellent agreement
between the theoretical prediction~(\ref{lyap5}) and the actual behaviour
of the Lyapunov exponent $\lambda$, as can be seen from Fig.~\ref{weaal0}
and Fig.~\ref{weaal1} which represent $\lambda$ as a function of the
energy $E$ for two values of the parameter $\alpha$.
% (Note: weaal0 = WEAk disorder, ALpha = 0)

\section{Strong (uncorrelated) disorder}
\label{sud}

\subsection{Analytical estimates for the localization length}

For the strong disorder case we consider the model~(\ref{ham}) with
diagonal terms of the form~(\ref{diadis}), i.e. $\delta_{n} = \alpha
\epsilon_{n}$ with $|\alpha| < 2$ . The Schr\"{o}dinger equation is then
given by~(\ref{schr4}) and the model is completely defined once
statistical properties of the random variables $\epsilon_{n}$ are
specified.
Throughout this Section we assume that the energies $\epsilon_{n}$ are
{\em independent} random variables with a common probability distribution
given by the box distribution~(\ref{boxdis}) of width $W$.
The case of strong disorder is then defined by the condition $W \gg 1$.

Again, by introducing the rescaled amplitudes~(\ref{phi}) and the binding
energies~(\ref{xi}), we can reduce the Schr\"{o}dinger equation~(\ref{schr4})
to the form~(\ref{schr2}) with only diagonal disorder. In the latter model the
site-energies~(\ref{xi}) take now the form 
\begin{displaymath}
\xi_{n} = \frac{E - \alpha \epsilon_{n}}{1+\epsilon_{n}}
\end{displaymath}
and are independent random variables whose distribution $\tilde{p} \left(
\xi_{n} \right)$ is related to the probability $p \left( \epsilon_{n} \right)$
of the original variables $\epsilon_{n}$ by
\begin{displaymath}
\tilde{p} \left( \xi_{n} \right) = \frac{ \left| \alpha + E \right|}{
{\left( \alpha + \xi_{n} \right)}^{2}} \, p \left( \frac{E - \xi_{n}}{
\alpha + \xi_{n}} \right) .
\end{displaymath}
In explicit form, when $p \left( \epsilon_{n} \right)$ is the Anderson
distribution~(\ref{boxdis}), the corresponding probability for the variables
$\xi_{n}$ reads
\begin{equation}
\tilde{p} \left( \xi_{n} \right) = \left\{
\begin{array}{ccc}
\frac{\left| \alpha + E \right|}{\left( \alpha + \xi_{n} \right)^{2}}
\frac{1}{W} & \mbox{ if } & \xi_{n} < \xi^{(1)} \mbox{ or } 
\xi^{(2)} < \xi_{n} \\
0 & \mbox{ if } & \xi^{(1)} < \xi_{n} < \xi^{(2)}
\end{array}
\right.
\label{noisdis}
\end{equation}
where
\begin{equation}
\xi^{(1)} = \min \left\{ \frac{2 E +\alpha W}{2 - W},
\frac{2 E - \alpha W}{2 + W}
\right\}
\label{xi1}
\end{equation}
and
\begin{equation}
\xi^{(2)} = \max \left\{ \frac{2 E + \alpha W}{2 - W},
\frac{2 E - \alpha W}{2 + W}
\right\} .
\label{xi2}
\end{equation}

Following the Hamiltonian map approach already used in the previous
Section, we can cast equation~(\ref{schr2}) in the form of the
two-dimensional map
\begin{equation}
\left( \begin{array}{c} \phi_{n+1} \\
                        \phi_{n}
       \end{array}
\right) =
{\bf T}_{n}
\left( \begin{array}{c} \phi_{n} \\
                        \phi_{n-1}
       \end{array}
\right),
\label{map3}
\end{equation}
where ${\bf T}_{n}$ is the transfer matrix
\begin{equation}
{\bf T}_{n} =
\left( \begin{array}{cc} \xi_{n} & -1 \\
                         1       &  0 
       \end{array}
\right) .
\label{transfer}
\end{equation}
By introducing polar coordinates through the relations
$\phi_{n} = r_{n} \cos \theta_{n}$ and $\phi_{n-1} = r_{n} \sin \theta_{n}$,
one can write the map~(\ref{map3}) in the form
\begin{equation}
\left\{ \begin{array}{ccl}
\cos \theta_{n+1} & = & \frac{1}{D_{n}} \sin \theta_{n} \\
\sin \theta_{n+1} & = & - \frac{1}{D_{n}} \left[ \cos \theta_{n}
+ \xi_{n} \sin \theta_{n} \right]
\end{array} \right.
\label{stromap}
\end{equation}
where
\begin{displaymath}
D_{n} = \frac{r_{n+1}}{r_{n}} = \sqrt{1 + \xi_{n} \sin \left( 2 \theta_{n}
\right) + \xi_{n}^{2} \sin^{2} \theta_{n} } .
\end{displaymath}
As before, the Lyapunov exponent is given by the relation~(\ref{lyap2}) and
can, therefore, be expressed by the integral
\begin{equation}
\lambda = \int  d\xi \, \tilde{p} \left( \xi \right) \int d\theta \,
\rho_{\xi} \left( \theta \right) \log D(\xi,\theta)
\label{lyap6}
\end{equation}
where the symbol $\rho_{\xi} \left( \theta \right)$ stands for the
invariant measure of the variable $\theta$ of the map~(\ref{stromap}), and
the subscript $\xi$ stands in order to stress that such a measure depends
on the noise distribution $\tilde{p} \left( \xi_{n} \right)$.

\subsection{Decomposition of the invariant measure}

\begin{figure}[htb]
\begin{center}
\epsfig{file=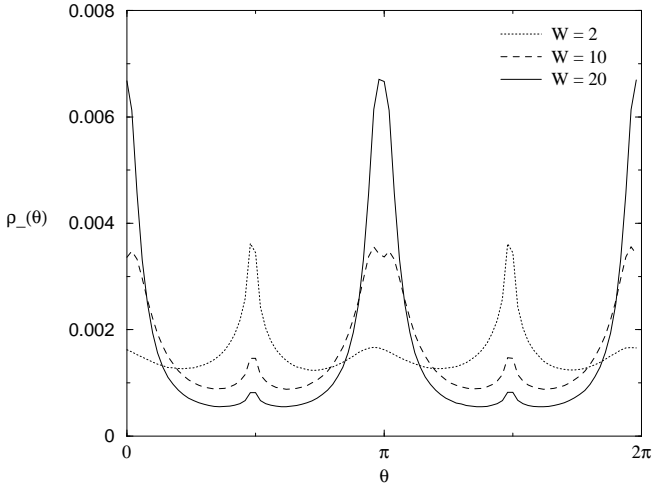,width=3.4in,height=2.6in}
\caption{Invariant measure $\rho_{-}\left( \theta \right)$ for various
noise strengths (data refer to the energy value E=0.01 and $\alpha=0$)}
\label{invmis}
\end{center}
\end{figure}

The integral~(\ref{lyap6}) can be separated in two parts,
$\lambda = I_{+} + I_{-}$
%\begin{displaymath}
%\lambda = I_{+} + I_{-}
%\end{displaymath}
with
\begin{displaymath}
I_{\pm} = \int  d\xi \, \tilde{p}_{\pm} \left( \xi \right) \int d\theta \,
\rho_{\pm} \left( \theta \right) \log D(\xi,\theta) .
\end{displaymath}
Here $\tilde{p}_{\pm} \left( \xi \right)$ are the noise distributions
\begin{displaymath}
\tilde{p}_{+} \left( \xi \right) = \left\{
\begin{array}{cc}
\tilde{p} \left( \xi \right) & \mbox{ for } \left| \xi \right| > 2 \\
0 & \mbox{ elsewhere}
\end{array}
\right.
\end{displaymath}
and
\begin{displaymath}
\tilde{p}_{-} \left( \xi \right) = \left\{
\begin{array}{cc}
\tilde{p} \left( \xi \right) & \mbox{ for } \left| \xi \right| < 2 \\
0 & \mbox{ elsewhere}
\end{array}
\right.
\end{displaymath}
while $\rho_{+} \left( \theta \right)$ and $\rho_{-} \left( \theta \right)$ 
are the corresponding invariant measures.
This decomposition is meaningful because numerical computations have shown
that, upon increasing the noise strength $W$, the invariant measure
$\rho_{-} \left( \theta \right)$ becomes more and more sharply
peaked around the single values $\theta = \pi k$ with $k$ integer
(as can be clearly seen from the Fig.~\ref{invmis}).
As a consequence, the integral $I_{-}$ becomes negligible for strong
disorder. The formula~(\ref{lyap6}) can therefore be approximated by
\begin{equation}
\lambda \simeq \int_{\left| \xi \right| > 2}  d\xi \, \tilde{p} \left( \xi
\right) \int d\theta \, \rho_{+} \left( \theta \right) \log D(\xi,\theta) .
\label{lyap7}
\end{equation}
The analysis of this expression reveals the simple dependence
$\lambda \propto 1/W$ of the Lyapunov exponent on the noise strength.
In fact, in (\ref{lyap7}) the $W$-dependence of the integral is
introduced in a twofold way, through the functional $1/W$ dependence of
the noise distribution~(\ref{noisdis}) and through the integration range for
the $\xi$-variable. For large enough $W$, the integration range is simply
determined by the condition $|\xi| > 2$ because the limits~(\ref{xi1})
and~(\ref{xi2}) of the interval where the distribution~(\ref{noisdis}) is
zero are located inside the interval $|\xi| < 2$. Therefore, for the
integral~(\ref{lyap7}) the only dependence left on the noise strength is the
one due to the factor $1/W$ contained in the integrand.
Physically, this means that, for strong disorder, the localization length
grows linearly with the noise intensity. In contrast to the standard
Anderson model, therefore, an increase of the lattice disorder reduces
the localization of the electronic wave-functions instead of enhancing it.

\subsection{Another approach for the computation of $\lambda$}
Although formula~(\ref{lyap7}) allows one to predict the relation of
inverse proportionality between $\lambda$ and $W$ without carrying out
any explicit calculations, the determination of the proportionality constant
requires a direct evaluation of the integral.
Unfortunately, the analytical computation of integral~(\ref{lyap7}) is out
of reach, since there is no way to determine the invariant measure
$\rho_{+}(\theta)$. One can overcome this difficulty using a different
approach successfully used in~\cite{Izr98} for an analogous problem.

Once more, we make use of the fact that the Schr\"{o}dinger
equation~(\ref{schr2}) is exactly equivalent to the 2D
map~(\ref{map3}) whose eigenvalues are
\begin{equation}
\lambda_{n}^{(1,2)} = \frac{\xi_{n} \pm \sqrt{\xi_{n}^{2} - 2}}{2} .
\label{eigval}
\end{equation}
For $|\xi_{n}| < 2$ the eigenvalues~(\ref{eigval}) take the form
$\lambda_{n}^{(1,2)} = e^{\pm i \mu_{n}}$ with $\xi_{n} = 2 \cos \mu_{n}$,
so that, at each step, the evolution dictated by the map~(\ref{map3}) results
in a simple rotation in the two-dimensional phase-space.
For $|\xi_{n}| > 2$, however, both eigenvalues of the map~(\ref{map3}) are
real and this consideration allows one to compute the Lyapunov
exponent~(\ref{lyap7}) using the approximated expression
\begin{equation}
\lambda = \langle \log \left| \Lambda \right| \rangle =
\int d\xi \, \tilde{p}_{+} \left( \xi \right) \log  \Lambda (\xi)
\label{lyap8}
\end{equation}
where
$\Lambda = \max \left\{ \left| \lambda^{(1)} \right|, \left| \lambda^{(2)}
\right| \right\}$.
%\begin{displaymath}
%\Lambda = \max \left\{ \left| \lambda^{(1)} \right|, \left| \lambda^{(2)}
%\right| \right\} .
%\end{displaymath}
The evaluation of integral~(\ref{lyap8}) leads to the formula
\begin{equation}
\lambda = \gamma \left( \alpha \right) \frac{ \left| E + \alpha \right|}{W}
\label{strolyap}
\end{equation}
where
\begin{displaymath}
\gamma \left( \alpha \right) =
2 \int_{2}^{\infty}  d\xi  \;\; \frac{\alpha^{2} + \xi^{2}}
{\left( \alpha^{2} - \xi^{2} \right)^{2}} \;\;
\log \left( \frac{\xi + \sqrt{ \xi^{2} - 4}}{2} \right) .
\end{displaymath}

\begin{figure}[htb]
\begin{center}
\epsfig{file=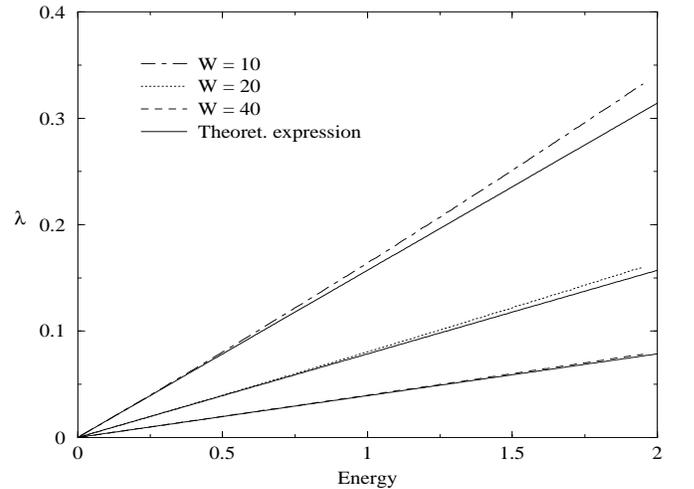,width=3.4in,height=2.6in}
\caption{Lyapunov exponent vs. energy for various noise strengths with
$\alpha = 0$}
\label{stral0}
\end{center}
\end{figure}

\begin{figure}[htb]
\begin{center}
\epsfig{file=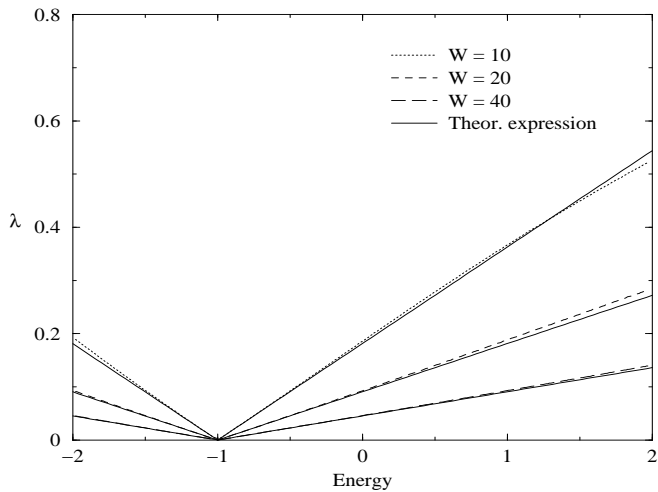,width=3.4in,height=2.6in}
\caption{Lyapunov exponent vs. energy for various noise strengths with
$\alpha = 1$}
\label{stral1}
\end{center}
\end{figure}

The validity of the formula~(\ref{strolyap}) was numerically checked; the
computations showed a good agreement between the theoretical prediction
and the actual behaviour of the Lyapunov exponent, as can be seen, for
instance, from Fig.~\ref{stral0} and Fig.~\ref{stral1}.
% (Note: stral0 = STRong disorder, ALpha=0)

Notice that the Lyapunov exponent~(\ref{strolyap}) exhibits the expected
$1/W$ dependence on the noise strength. As we explained, this is a
consequence of the fact that the region $\xi^{(1)} < \xi < \xi^{(2)}$ where
the integrand of formula~(\ref{lyap8}) is zero lies {\em outside} the
integration domain $|\xi| > 2$. In passing we remark that, if the condition
$|\alpha|<2$ were not fulfilled, this conclusion would not be true. Therefore,
for $|\alpha| > 2$ the integral~(\ref{lyap8}) would depend on $W$ not only
through the integrand but also through the limits of the integration
domain. The computation of the integral remains possible
also in this case, but then the result is no longer as simple as the one
expressed by formula~(\ref{strolyap}). That is why we chose to
restrict the range of the parameter $\alpha$ to the interval $|\alpha|<2$.

\section{Disorder with short-range correlations}
\label{dwsrc}

In this Section we focus our attention to the case in which the diagonal
and off-diagonal disorder in the general model~(\ref{ham}) exhibit
short-range correlations.
More precisely, we consider the model defined by assuming that the
couples of random variables $\zeta_{n} = \left( \delta_{n}, \epsilon_{n}
\right)$ in the Schr\"{o}dinger equation~(\ref{schr1}) fulfill the
following conditions:
\begin{enumerate}
\item The energies $\zeta_{n}$ can take only two values: $\zeta_{+} =
\left( \delta_{+}, \epsilon_{+} \right)$ or $\zeta_{-} = \left( \delta_{-},
\epsilon_{-} \right)$.
\item In every succession $\left\{ \zeta_{n} \right\}$ the values
$\zeta_{+}$ and $\zeta_{-}$ appear in $N$-uples made of $N$ consecutive
identical terms.
\item The $N$-uples $\left( \zeta_{+}, \ldots ,\zeta_{+} \right)$ and
$\left( \zeta_{-}, \ldots ,\zeta_{-} \right)$ appear at random with
equal frequency along the succession $\left\{ \zeta_{n} \right\}$.
\end{enumerate}
If $N=2$, the pairs $\left( \zeta_{+},\zeta_{+} \right)$ and $\left(
\zeta_{-},\zeta_{-} \right)$ are called dimers; for $N=3$ the corresponding
triplets are named trimers, while for generic $N$ one speaks of `$N$-mers'.

Making use of the amplitudes~(\ref{phi}) and the energies~(\ref{xi}) as in
the previous cases, one can write the Schr\"{o}dinger equation~(\ref{schr1})
in the form~(\ref{schr2}) where the diagonal energies~(\ref{xi}) now take
two values
\begin{equation}
\xi_{\pm} = \frac{E - \delta_{\pm}}{1 + \epsilon_{\pm}}
\label{xipm}
\end{equation}
and appear in equiprobable $N$-uples randomly positioned along the
succession $\left\{ \xi_{n} \right\}$.
In other words, the model~(\ref{ham}), with both diagonal and off-diagonal
disorder specified by the preceding 1), 2), and 3) conditions, can be
associated with the $N$-mer model already studied in the literature
(see~\cite{Izr95} and references quoted therein).

As in reference~\cite{Izr95}, we can use the equivalence of the
model~(\ref{schr2}) with the map~(\ref{map3}) and study the solutions
of the Schr\"{o}dinger equation~(\ref{schr2}) in terms of the
trajectories of the map~(\ref{map3}).
With this interpretation in mind, we can address the problem of extended
states for the model~(\ref{schr2}) by considering a single $N$-mer of type
$\zeta_{+}$ embedded in an infinite chain of $\zeta_{-}$ $N$-mers.
In this case the orbits of the map~(\ref{map3}) are not affected by the
single $\zeta_{+}$ $N$-mer provided that the total transfer matrix through
the $N$-mer is equal to the identity matrix (apart from a plus/minus factor).
In other words, one must have
\begin{equation}
{{\bf T}_{+}}^{N} = \pm {\bf E}
\label{transp}
\end{equation}
where ${\bf T_{+}}$ is the transfer matrix~(\ref{transfer}) (with
$\xi_{n} = \xi_{+}$) and ${\bf E}$ is the unit matrix.

The `transparency' condition~(\ref{transp}) can be satisfied provided
that the stability condition $\left| \xi_{+} \right| < 2$ for the
map~(\ref{map3}) is fulfilled.
In this case the eigenvalues of the matrix ${\bf T}_{+}$ have the form
$e^{\pm i \mu_{+}}$ (with $\xi_{+} =2 \cos \mu_{+}$) and the
equation~(\ref{transp}) is equivalent to $\exp (i \mu_{+} N) = \pm 1$
%\begin{displaymath}
%e^{i \mu_{+} N} = \pm 1
%\end{displaymath}
which in turn leads to
$\mu_{+} = \pi k / N$ with $k$ integer.
%\begin{displaymath}
%\mu_{+} = \frac{\pi}{N} k \;\;\;\; \mbox{ with $k$ integer}.
%\end{displaymath}
On the other hand, the stability condition $\left| \xi_{+} \right| < 2$
requires to discard the $k$ values
$k = m \cdot N$ with $m$ integer.
%\begin{displaymath}
%k = m \cdot N \;\;\;\; \mbox{ with $m$ integer}.
%\end{displaymath}
Using~(\ref{xipm}) and the relation $\xi_{+} = 2 \cos \mu_{+}$, we come
to the conclusion that the $N$-mer is transparent for the following
energy values
\begin{displaymath}
E = \delta_{+} + 2 \left( 1 + \epsilon_{+} \right) \cos \left(
 \frac{\pi}{N} k \right)
\end{displaymath}
with $k=1,2,\ldots,N-1$.

The study of the single $N$-mer case allows one to understand that, in the
general case of a chain where $N$-mers of type $\zeta_{+}$ and $\zeta_{-}$
randomly alternate, the model~(\ref{ham}) with disorder correlated by
$N$-uples has extended states when the energy takes the values
\begin{equation}
E = \left\{ \begin{array}{c}
           \delta_{+} + 2 \left( 1 + \epsilon_{+} \right)
           \cos \left( \frac{\pi k}{N} \right) \\
           \delta_{-} + 2 \left( 1 + \epsilon_{-} \right)
           \cos \left( \frac{\pi k}{N} \right)
           \end{array}
\right.
\label{enetra}
\end{equation}
where $k = 1,2,\ldots,N-1$.
In the particular case when the diagonal and off-diagonal disorder are
related by the linear equation $\delta_{n} = \alpha \epsilon_{n}$, one
should add the special value $E = - \alpha$ to the list~(\ref{enetra}),
because in this case the variables~(\ref{xipm}) take the single value
$\xi_{n} = - \alpha$ and disorder disappears from equation~(\ref{schr2}).

\begin{figure}[htb]
\begin{center}
\epsfig{file=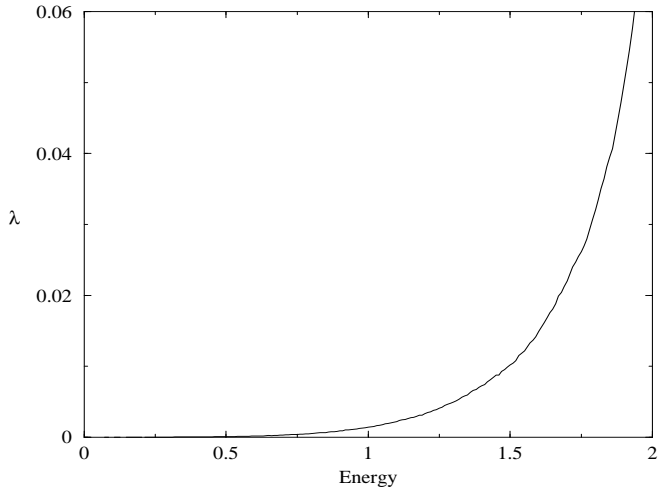,width=3.4in,height=2.6in}
\caption{Lyapunov exponent for dimer-correlated disorder with
$\delta_{\pm} = 0$, $\epsilon_{+}=0$, $\epsilon_{-} = 0.5$}
\label{dimer}
\end{center}
\end{figure}

\begin{figure}[htb]
\begin{center}
\epsfig{file=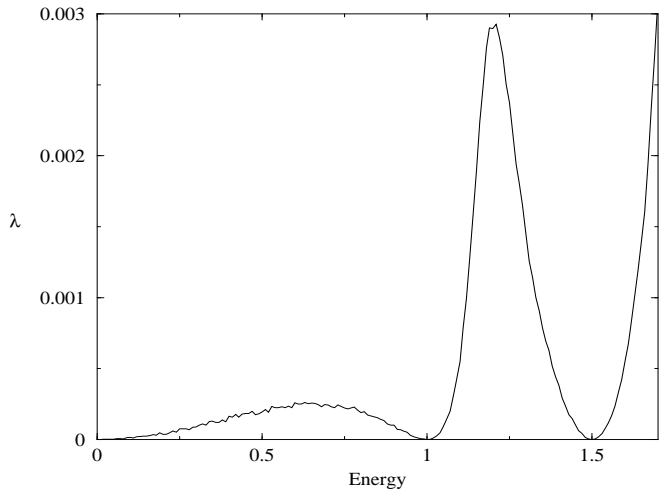,width=3.4in,height=2.6in}
\caption{Lyapunov exponent for trimer-correlated disorder with
$\delta_{\pm} = 0$, $\epsilon_{+} = 0$, $\epsilon_{-} = 0.5$}
\label{trimer}
\end{center}
\end{figure}

The theoretical prediction~(\ref{enetra}) was checked by numerically
computing the Lyapunov exponent for the system~(\ref{ham}) with dimer-
and trimer-correlated disorder as a function of the energy. Numerical
experiments have confirmed the validity of formula~(\ref{enetra}),
specifically, the Lyapunov exponent vanishes whenever the energy takes one
of the values~(\ref{enetra}), see Figs.~\ref{dimer},~\ref{trimer}.
% (Note: DIMER correlated disorder)
% (Note: TRIMER correlated disorder)

\section{Concluding remarks}
\label{cr}

In this paper we have studied the 1D tight-binding model described by
the Hamiltonian~(\ref{ham}) which has both diagonal and off-diagonal
variable matrix elements. The main object of this study is the analysis
of the localization length and its dependence on underlying correlations
in the diagonal and off-diagonal potential.
By making use of the reduction of the quantum model to a proper classical
Hamiltonian map, we have considered a few most interesting cases of the
interplay between diagonal and off-diagonal correlated disorder.

The first case which has been analysed in details is the model with
a weak site-potential and a weak fluctuating part in the off-diagonal matrix
elements. In this case we have derived a general analytical expression for
the localization length valid for any kind of randomness in the potential.
More specifically, the localization length is expressed in terms of pair
correlators for the diagonal and off-diagonal matrix elements.
Using this expression, one can consider any kind of correlated disorder,
as well as deterministic potential and hopping off-diagonal amplitudes.
The expression obtained for the localization length allows to reveal how
this latter quantity is influenced by pair correlations. A few specific
cases of correlated disorder have been discussed to illustrate the role
of the correlations.
One of the most interesting results is that if the diagonal and
off-diagonal disorder are proportional to each other, there is a
specific value of the energy for which the eigenstates are extended. For
finite sample, therefore, there is a region where all states are fully
transparent as in the dimer models with short-range correlations.

Another case studied in this paper is that of strong disorder (both
diagonal and off-diagonal). In this case we have been able to obtain an
approximate analytical expression for the localization length which
shows that, when lattice randomness increases, localization effects
weaken. This feature contrasts with the behaviour of the standard 1D
Anderson model (with only diagonal disorder) and is a non-trivial
consequence of the interaction of the site-potential with the off-diagonal
hopping energies.

Very recently \cite{OKG99} an analytical expression for the invariant
density $\rho_{\xi}(\theta)$ in the strong disorder case has been obtained
in a different context.

Finally we have examined the case of disorder with short-range correlation,
focusing our attention on the specific $N$-mer model, where disorder
correlations are introduced by assigning the same value of the site- and
hopping-energies to $N$ consecutive chain sites. For this case we have
showed how correlations give rise to a discrete set of extended states.

\section{Acknowledgments}

The authors are thankful to A.Krokhin for discussions and valuable comments.  
FMI gratefully acknowledges support by the CONACyT (Mexico) Grants
No. 26163-E and No. 28626-E.

\end{multicols}
\end{document}